\pdfoutput=1
\documentclass[12pt,a4paper]{iopart}

\usepackage{iopams,graphicx}
\usepackage{cite}

\usepackage{bm}% bold math
\newcommand{\tmmathbf}[1]{{\boldsymbol{#1}}}
\newcommand{\mathbbm}{\mathbb}

\begin{document}

\title{Wigner function for the orientation state}

\author{Timo Fischer}
\address{University of Duisburg-Essen, Lotharstra{\ss}e 1-21, 47048 Duisburg, Germany}
\author{Clemens Gneiting}
\address{Albert-Ludwigs University of Freiburg,
Hermann-Herder-Stra{\ss}e 3, 79104 Freiburg, Germany }
\author{Klaus Hornberger}
\address{University of Duisburg-Essen, Lotharstra{\ss}e 1-21, 47048 Duisburg, Germany}

\begin{abstract}
We introduce a quantum phase space representation for the orientation state of extended quantum objects, using the Euler angles and their conjugate momenta as phase space coordinates. It exhibits the same properties as the standard Wigner function and thus provides an intuitive framework for discussing quantum effects and semiclassical approximations in the rotational motion. Examples illustrating the viability of this quasi-probability distribution include the phase space description of a molecular alignment effect.
\vspace*{2\baselineskip}

\noindent
Published in: {\sf New J. Phys. 15, 06004 (2013) }
\end{abstract}

%

%02.20.Qs     General properties, structure, and representation of Lie groups
%03.65.-w     Quantum mechanics
%03.65.Ca     Formalism
%05.30.-d     Quantum statistical mechanics
%33.80.-b     Photon interactions with molecules (see also 42.50.-p Quantum optics)
%37.90.+j     Other topics in mechanical control of atoms, molecules, and ions (restricted to new topics in section 37)
%42.50.-p     Quantum optics (for lasers, see 42.55.-f and 42.60.-v; see also 42.65.-k Nonlinear optics; 03.65.-w Quantum mechanics)
%42.50.Ct     Quantum description of interaction of light and matter; related experiments
%03.65.Sq Semiclassical theories in quantum mechanics, 

\maketitle

\section{Introduction}

Soon after Heisenberg and Schr\"odinger invented modern quantum mechanics, Wigner introduced his prescription to represent quantum states by distribution functions in classical phase space {\cite{EPW32}}. 
Moyal's discovery {\cite{JEM49}} that this mapping between quantum operators and phase space functions reflects Weyl's correspondence rule {\cite{HW28}} led to a representation of quantum mechanics as a statistical theory on classical phase space. The Wigner function has found numerous applications in many areas of physics, ranging from solid state physics, optics, quantum optics and interferometry with particles to collisions of molecules and chemical reactions {\cite{WRF87,MJB79,UL95,CK97,HL80,HW98,WS01}}.

Wigner's standard formulation is restricted to the center-of-mass motion of spinless particles. Many extensions to other dynamical degrees of freedom have been suggested, notably by means of mappings to the Cartesian case \cite{Pierre1969251,Ni701,Ni702}, or formulations based on Lie groups 
\cite{GSA81,JCV89,HF90,JPD94,CB98,CB99,Na99,NM04,NM05,KR08}.
The former comes with major drawbacks; for instance, one cannot even calculate expectation values for arbritrary operators in these frameworks. In \cite{GSA81,JCV89,JPD94} a phase space formulation for single irreducible representations of SU(2) is introduced, i.e.\ for spin states with a fixed value of $j$. In \cite{KR08} this is extended to account for superpositions in $j$. Based on the theory of generalized coherent states \cite{AP86} these concepts are applied to quantum systems possessing arbitrary Lie group symmetries in \cite{HF90,CB98,CB99}. However, in all these approaches the marginal distributions cannot be obtained from the Wigner function by integrating out the other variables.
In \cite{NM04,NM05} a phase space formulation is developed for quantum systems whose \emph{configuration space} is a Lie group. This work takes into consideration the desired features of the standard Wigner function. But also here the marginal distributions cannot be calculated as one would do with a probability distribution, thus impeding a quasi-probability interpretation. All the essential features of the standard Wigner function have so far only been demonstrated for the simple one-dimensional case of a single angle-angular momentum pair \cite{mukunda1978,NM79,Leonhardt1995,
leonhardt1996discrete,Rigas11}.
Given the success of the Wigner-Weyl representation for point particles, a viable phase space description of spatially extended quantum objects  should be useful e.g.~for experiments probing quantum and classical dynamics in the rotation and alignment of complex molecules. 

In this article we present a Wigner function for the orientation state of an extended quantum object. In contrast to the approaches mentioned above, it is founded on the quantized canonical coordinates of the underlying classical phase space, providing a one-to-one mapping between quantum operators and classical phase space functions, the Weyl symbols. In close analogy to the standard Moyal formulation, we find that any Weyl-ordered observable is mapped to its equivalent function on phase space, that integrating out phase space coordinates yields the reduced probability distribution, and that the motion of the Wigner function is described by a quantum Liouville equation, which turns into its classical equivalent as $\hbar \rightarrow 0$.

\section{Wigner function for the orientation}

The Hilbert space
$\mathcal{H}$ of a rigid rotor is spanned by the vectors $|JKM \rangle$, the
eigenstates of the square of the angular momentum operator $\mathsf{J}^2$, and
its projections $\mathsf{P} _z$ and $\mathsf{J} _z$ on the body-fixed and
space-fixed $z$-axis. An alternative basis is provided by the resolution of
the identity 
\begin{equation}
\mathsf{} \mathbbm{I} = \int_0^{2 \pi} \mathrm{d} \alpha
\int_0^{\pi} \mathrm{d} \beta \sin \beta \int_0^{2 \pi} \mathrm{d} \gamma \,  |
\alpha, \beta, \gamma \rangle \langle \alpha, \beta, \gamma |\end{equation}
in terms of the
Euler angles $\left( \alpha, \beta, \gamma \right) \equiv \Omega$, which serve
to specify the orientation (in the usual $z$-$y$-$z$ convention). The basis
transformation 
\begin{equation}
\langle \Omega |JKM \rangle = \sqrt{\frac{2 J + 1}{8 \pi^2}}
D^{J \ast}_{MK} (\Omega)
\end{equation}
is mediated by the Wigner D-matrices $D^J_{MK}
\left( \Omega \right)$ {\cite{BieLou84}}.

Choosing the Euler angles as natural configuration space coordinates, the
classical phase space is completed by their conjugate momenta $p_{\alpha}, p_{\beta}, p_{\gamma}$. As the generators of the rotations defining the Euler
angles they are projections of the angular momentum vector
${\boldsymbol{J}}$ onto the corresponding rotation axes: \
$p_{\alpha}$ is the projection of ${\boldsymbol{J}}$ on the space-fixed $z$-axis, $p_{\beta}$ on the nodal line formed by intersecting the space-fixed $x$-$y$ plane with the body-fixed $x$-$y$ plane, and $p_{\gamma}$ on the
body-fixed $z$-axis.

Quantum mechanically, the symmetrized projections are given by the operators  \cite{BieLou84}
\numparts
\begin{eqnarray}
\mathsf{p}_{\alpha} &=& - i \hbar \partial_{\alpha} \\ \mathsf{p}_{\beta} &=& - i
\hbar \left( \partial_{\beta} + \frac{1}{2} \cot \beta \right)
\\
\mathsf{p}_{\gamma} &=& - i \hbar \partial_{\gamma}\,. 
\end{eqnarray}
\endnumparts
Like $\mathsf{J}^2$,
$\mathsf{J}_z$, $\mathsf{P}_z$, they form a complete set of commuting
observables exhibiting a discrete spectrum, with eigenvalues
\begin{equation}
p_{\alpha} =
\hbar m_{\alpha}, \quad p_{\beta} = 2 \hbar m_{\beta} \quad\mathrm{and}\quad  
p_{\gamma} = \hbar m_{\gamma},\quad  m_{\alpha, \beta, \gamma} \in \mathbbm{Z}.
\end{equation}We denote the
corresponding eigenstates as $|m_{\alpha} m_{\beta} m_{\gamma} \rangle \equiv
|{\boldsymbol{m}} \rangle$; they provide a further basis of
$\mathcal{H}$ and their Euler angle representation is \begin{equation}
\langle \Omega
|{\boldsymbol{m}} \rangle = \frac{1}{\sqrt{4 \pi^3 \sin \beta}}\,e^{im_{\alpha} \alpha} e^{i 2 m_{\beta}
\beta} e^{im_{\gamma} \gamma} 
\end{equation}

With this, we can now state our Wigner function 
\begin{equation}
W \left( \Omega,
{\boldsymbol{m}} \right) \equiv  W \left( \alpha, \beta, \gamma,
\frac{p_{\alpha}}{ \hbar}, \frac{p_{\beta}}{ 2 \hbar}, \frac{p_{\gamma}}{ \hbar} \right)
\end{equation}
in terms
of both the orientation representation $\langle \Omega' | \rho | \Omega''
\rangle$ and the angular momentum representation $\langle
{\boldsymbol{m}}' | \rho |{\boldsymbol{m}}'' \rangle$ of
the quantum state, 
\numparts
\begin{eqnarray}
\fl
  W \left( \Omega, {\boldsymbol{m}} \right)  &= & \frac{1}{4 \pi^3}
  \int_{- \pi}^{\pi} \mathrm{d} \alpha'  \int^{\pi / 2}_{- \pi / 2} \mathrm{d}
  \beta' \int^{\pi}_{- \pi} \mathrm{d} \gamma'  \sqrt{\sin \beta'_+ \sin
  \beta'_-} \,e^{im_{\alpha} \alpha'} e^{i 2 m_{\beta} \beta'} e^{im_{\gamma}
  \gamma'} \nonumber
  \\
  && \times 
  \langle \alpha'_-, \beta'_-, \gamma'_- | \rho | \alpha'_+,
  \beta'_+, \gamma'_+ \rangle  \label{eq:wignerExplicitP}\\
  & = & \frac{1}{4 \pi^3} 
  \sum_{\tmmathbf{m}', \tmmathbf{m}'' \in \mathbbm{Z}^3}
  {\mathrm{sinc}} \left[ \left(
  {\boldsymbol{m}}- \frac{{\boldsymbol{m}}'
  +{\boldsymbol{m}}''}{2} \right) \pi \right] e^{i \left(
  m'_{\alpha} - m_{\alpha}'' \right) \alpha} e^{i 2 \left( m'_{\beta} -
  m''_{\beta} \right) \beta}
  \nonumber\\
  &&\times e^{i \left( m'_{\gamma} - m''_{\gamma} \right)
  \gamma}   \langle {\boldsymbol{m}}' | \rho
  |{\boldsymbol{m}}'' \rangle \,.  \label{eq:wignerExplicitM}
\end{eqnarray}
\endnumparts
Here we use the abbreviations $\alpha'_{\pm} = \left( \alpha \pm
\alpha' / 2 \right) {\mathrm{mod}}2 \pi$, alike for $\gamma$,
$\beta'_{\pm} = \left( \beta \pm \beta' / 2 \right)
{\mathrm{mod}} \pi$, and ${\mathrm{sinc}}
\left( {\boldsymbol{x}} \right) = \prod_i \sin \left( x_i \right) /
x_i$. While the angular representation (\ref{eq:wignerExplicitP}) is similar to the standard Wigner function {\cite{EPW32}}, the momentum representation (\ref{eq:wignerExplicitM}) differs notably exhibiting a double sum and sinc functions, which act like blurred Kronecker deltas. 
The Wigner function in terms of the symmetric top eigenstate basis $\langle JKM| \rho |J' K' M' \rangle$ is readily obtained from (\ref{eq:wignerExplicitP}) by inserting the corresponding resolution of the identity.

Note that the angular momentum arguments ${\boldsymbol{m}}$ are integers, which reflects the discrete spectra of the corresponding operators. As discussed below, this discontinuity in the momenta poses no conceptual problem, as all relevant properties of the standard Wigner function remain untouched.  The discreteness is rather a necessary consequence of quantizing compact coordinates, which disappears in the classical limit. 

As in the standard case, one can introduce a Stratonovich-Weyl operator kernel
$\Delta \left( \Omega, {\boldsymbol{m}} \right)$ (see below) {\cite{KEC69}},
such that the mapping of any operator $\mathsf{A}$ onto its Weyl symbol
$W_{\mathsf{A}} \left( \Omega, {\boldsymbol{m}} \right)$ reads as
\begin{eqnarray}
  W_{\mathsf{A}} \left( \Omega, {\boldsymbol{m}} \right) & = &
  {\mathrm{tr}} \left[ \mathsf{A} \Delta \left( \Omega,
  {\boldsymbol{m}} \right) \right]  \label{eq:weylsymbol}
\end{eqnarray}
and its inversion
\begin{eqnarray}
  \mathsf{A} & = & \frac{1}{4 \pi^3} \sum_{{\boldsymbol{m}}}
  \int_0^{2 \pi} \mathrm{d} \alpha \int_0^{\pi} \mathrm{d} \beta \int_0^{2
  \pi} \mathrm{d}  \gamma \,W_{\mathsf{A}}  \left( \Omega,
  {\boldsymbol{m}} \right) \Delta \left( \Omega,
  {\boldsymbol{m}} \right) . \label{eq:inversewigner} \nonumber
\end{eqnarray}
For $\mathsf{A} = \rho$ this implies (\ref{eq:wignerExplicitP}) and (\ref{eq:wignerExplicitM}), since $W_{\rho} \left(
\Omega, {\boldsymbol{m}} \right) = 4 \pi^3 W \left( \Omega,
{\boldsymbol{m}} \right)$.

\section{Relevant properties of the Wigner function}Remarkably, all
relevant properties of the standard Wigner function can be recovered. It
follows from (\ref{eq:wignerExplicitP}) that the Weyl symbol of any hermitian
operator is real, as required for a proper phase space representation.
Moreover, $W \left( \Omega, {\boldsymbol{m}} \right)$ is normalized
with respect to the phase space integral, 
\begin{equation}
\int \mathrm{d} \Omega
\sum_{{\boldsymbol{m}}} W \left( \Omega,
{\boldsymbol{m}} \right) = 1 \,,
\end{equation}
where $\mathrm{d} \Omega \equiv
\mathrm{d} \alpha \mathrm{d} \beta \mathrm{d} \gamma$, and expectation values
can be calculated by the phase space average 
\begin{equation}
\int \mathrm{d} \Omega
\sum_{{\boldsymbol{m}}}  W_{\mathsf{A}} \left( \Omega,
{\boldsymbol{m}} \right) W \left( \Omega,
{\boldsymbol{m}} \right)  = \langle \mathsf{A} \rangle\,.
\end{equation}
The
quasi-probability interpretation of the Wigner function manifests itself in
the expressions for the probability distributions $\langle \Omega | \rho |
\Omega \rangle$ and $\langle {\boldsymbol{m}}| \rho
|{\boldsymbol{m}} \rangle$. They can be obtained as marginal
distributions of $W \left( \Omega, {\boldsymbol{m}} \right)$ by
integrating out the conjugate variable,
\numparts
\begin{eqnarray}\label{eq:p1}
\sum_{{\boldsymbol{m}}} W
\left( \Omega, {\boldsymbol{m}} \right) = \sin( \beta )\langle \Omega
| \rho | \Omega \rangle
\\\label{eq:p2}
\int W \left( \Omega,
{\boldsymbol{m}} \right) \mathrm{d} \Omega = \langle
{\boldsymbol{m}}| \rho |{\boldsymbol{m}} \rangle
\end{eqnarray}
\endnumparts
as is readily verified from (\ref{eq:wignerExplicitP}) and
(\ref{eq:wignerExplicitM}).

As for the Weyl symbols of the coordinate and momentum operators,
Eq.~(\ref{eq:wignerExplicitP}) yields the expected expressions
$W_{\hat{\alpha}} \left( \Omega, {\boldsymbol{m}} \right) = \alpha$
and $W_{\mathsf{p}_{\alpha}} \left( \Omega, {\boldsymbol{m}}
\right) = \hbar m_{\alpha} = p_{\alpha}$. Moreover, in close analogy to the
standard case, the operator ordering 
\begin{equation}\label{eq:p3}
\left\{ \hat{\alpha}^n,
\mathsf{p}^m_{\alpha} \right\}_W = \frac{1}{2^{ m}} \sum_{k = 0}^m
{m \choose k}
\mathsf{p}_{\alpha}^{m - k}  \hat{\alpha}^n 
\mathsf{p}_{\alpha}^k
\end{equation}
turns the Weyl symbol of arbitrary products of
$\hat{\alpha}$ and $\mathsf{p}_{\alpha}$ into their classical equivalents,
\begin{equation}
W_{\left\{ \hat{\alpha}^n, \mathsf{p}^m_{\alpha} \right\}_W} \left( \Omega,
{\boldsymbol{m}} \right) = \alpha^n p_{\alpha}^m  \,;
\end{equation}
the other angles satisfy analogous expressions. Unlike in the standard case, this `Weyl ordering' is not equivalent to the `symmetric ordering' due to the operator-valued commutator $[\hat{\alpha}, \mathsf{p}_{\alpha} ]$.

We have thus established that  using the Euler angles and their conjugate momenta one can obtain  the desired features of a Wigner-Weyl phase space representation.  The phase space variables have a clear physical meaning and provide the appropriate framework for a semiclassical description of the quantum dynamics.

In view of the standard Wigner function one might expect that 
the phase space is continuous in both variables.
To appreciate why the momenta occur as discrete variables in the Wigner function, note that the momenta correspond to projections of the angular momentum vector. As such they are quantum mechanical observables with discrete spectra. This discreteness, which can in principle be observed experimentally, is a 
consequence of the compact range of the Euler angles.
Moreover, the eigenstates of these conjugate momenta are proper physical states. 
These two facts imply that the associated phase space coordinate of the Wigner function must be discrete, as can be seen from the fact that the Wigner function describing a mixture of such eigenstates must yield the correct marginal distribution for measurements of the conjugate momenta: after integrating out the angles it must yield a probability distribution for a discrete variable since the measurement outcomes are discrete.

The discreteness is thus a natural and unavoidable physical consequence of a consistent Wigner-Weyl representation, simply locating how closely one can get to a classical description. Outside the deep quantum regime, once the continuum limit can be taken, one recovers the full correspondence with classical mechanics, including the Liouville equation to leading order in $\hbar$.

\section{Construction of the Wigner function}To shed light on how
the Wigner function is devised, let us consider the displacement operators
\numparts
\begin{eqnarray}
\mathsf{D}_{\alpha} \left( \alpha, m_{\alpha} \right) = \exp \left(
im_{\alpha}  \hat{\alpha} \right) \exp \left( -\frac{ i \alpha \mathsf{p}_{\alpha} }{
\hbar}  \right)\,,
\\
\mathsf{D}_{\beta} \left( \beta,
m_{\beta} \right) = \exp ( i 2 m_{\beta}  \hat{\beta} ) \exp \left(
- \frac{i \beta \mathsf{p}_{\beta} }{ \hbar }\right)\,,
\\
\mathsf{D}_{\gamma} \left( \gamma, m_{\gamma} \right) = \exp \left(
im_{\gamma}  \hat{\gamma} \right) \exp \left( - \frac{ i \gamma \mathsf{p}_{\gamma}}{\hbar }\right)\,,
\end{eqnarray}
\endnumparts
where angular values outside the
Euler range are mapped back by taking the modulus, e.g.~$\mathsf{D}_{\alpha}
\left( \alpha, 0 \right) | \alpha_0 \rangle = | \left( \alpha_0 + \alpha
\right) {\mathrm{mod}}2 \pi \rangle$. These
phase space translation operators commute, as follows from the commutators of the phase space coordinates. The $\mathsf{D}
\left( \Omega, {\boldsymbol{m}} \right) = \mathsf{D}_{\alpha}
\left( \alpha, m_{\alpha} \right) \otimes \mathsf{D}_{\beta} \left( \beta,
m_{\beta} \right) \otimes \mathsf{D}_{\gamma} \left( \gamma, m_{\gamma}
\right)$ allow one to construct the major building block of the phase space
formulation, the operator kernel $\Delta \left( \Omega,
{\boldsymbol{m}} \right) = \mathsf{D} \left( \Omega,
{\boldsymbol{m}} \right) \Delta \left( 0, 0 \right)
\mathsf{D}^{\dag} \left( \Omega, {\boldsymbol{m}} \right)$, where
$\Delta \left( 0, 0 \right)$ is the direct product of
\numparts
\begin{eqnarray}
  \Delta_{\alpha} \left( 0, 0 \right) & = & \frac{1}{2 \pi} \sum_{m \in
  \mathbbm{Z}} \int^{\pi}_{- \pi} \mathrm{d} \alpha'   \, \mathsf{D}_{\alpha}
  \left( \alpha', m \right) e^{- im \alpha' / 2},  \label{eq:SWOKa}\\
  \Delta_{\beta} \left( 0, 0 \right) & = & \frac{1}{\pi} \sum_{m \in
  \mathbbm{Z}} \int^{\frac{\pi}{2}}_{- \frac{\pi}{2}} \mathrm{d} \beta'   \,
  \mathsf{D}_{\beta} \left( \beta', m \right) e^{- im \beta'}, 
  \label{eq:SWOKb}
\end{eqnarray}
\endnumparts
and $\Delta_{\gamma} \left( 0, 0 \right)$ having the same form as
(\ref{eq:SWOKa}). The choice of the phase factor in (\ref{eq:SWOKa}),
(\ref{eq:SWOKb}) guarantees that the angular momentum symmetry of a state is
correctly reflected by the Wigner function {\cite{Rigas11}}. The symmetric
choice of the integral limits, on the other hand, is required to ensure the hermiticity of the $\Delta_i \left( 0, 0 \right)$. Note that the negative
lower integration boundaries do not contradict the definition range of the Euler angles since they refer to translations.
The expressions (\ref{eq:wignerExplicitP}) and (\ref{eq:wignerExplicitM}) then follow by inserting into 
(\ref{eq:weylsymbol}) the explicit form
\begin{eqnarray}
\fl
  \Delta \left( \Omega,{\boldsymbol{m}} \right)  &= & 
  \int_{- \pi}^{\pi} \mathrm{d} \alpha'  \int^{\pi / 2}_{- \pi / 2} \mathrm{d}
  \beta' \int^{\pi}_{- \pi} \mathrm{d} \gamma'  \sqrt{\sin \beta'_+ \sin
  \beta'_-}\, e^{im_{\alpha} \alpha'} e^{i 2 m_{\beta} \beta'} e^{im_{\gamma}
  \gamma'} \nonumber
  \\
  && \times 
  | \alpha'_+,  \beta'_+, \gamma'_+ \rangle  
  \langle \alpha'_-, \beta'_-, \gamma'_- | \,.
\end{eqnarray}

These results reproduce a further crucial property of the Wigner function: it
properly reflects phase space translations in the sense that the Weyl symbol
of a translated operator $\mathsf{A}' = \mathsf{D}^{} \left( \Omega',
{\boldsymbol{m}}' \right) \mathsf{A}  \mathsf{D^{\dag}} \left(
\Omega', {\boldsymbol{m}}' \right)$ is the translated Weyl symbol
$W_{\mathsf{A}} \left( \Omega - \Omega',
{\boldsymbol{m}}-{\boldsymbol{m}}' \right)$. This is
proven by rewriting ${\mathrm{tr}} \left[ \mathsf{A}' \Delta
\left( \Omega, {\boldsymbol{m}} \right) \right]$ as
${\mathrm{tr}} \left[ \mathsf{A} \Delta \left( \Omega -
\Omega', {\boldsymbol{m}}-{\boldsymbol{m}}' \right)
\right]$ using the invariance of the trace under cyclic permutations.

\section{Quantum Liouville equation of a free symmetric
top}Applying the definition (\ref{eq:wignerExplicitP}) to the von
Neumann equation allows one to determine the quantum Liouville equation, the
law of motion for the Wigner function. To this end, the free symmetric top
Hamiltonian {\cite{TS05}} must be recast in terms of the canonical operators,
\begin{equation}
\mathsf{H} = \frac{( \mathsf{p}_{\alpha} - \mathsf{p}_{\gamma} \cos \hat{\beta} )^2 }{ ( 2 I_1 \sin^2 \hat{\beta} )} + \frac{\mathsf{p}_{\beta}^2}{ 2
I_1 }+ \frac{\mathsf{p}_{\gamma}^2}{ 2 I_3 }- \frac{\hbar^2 ( 1 + \sin^{- 2} \hat{\beta}) }{ 8 I_1 } \,,
\end{equation}
with $I_1$ and $I_3$ the principal moments of inertia. One
thus recovers the classical Hamiltonian with an additional quantum potential
given by the last term. Owing to this quantum
correction one does not recover the classical Liouville equation for the torque-free
motion, unlike the case of a free point particle. This is a consequence of
quantization and not inherent to the chosen phase space approach. Moreover,
the discreteness of the conjugate momenta yields summations over
${\boldsymbol{m}}$, in variance with the form of the classical
Liouville equation. However, in the limit that the Wigner function varies only
weakly in the discrete momenta they can be replaced by continuous variables.
This is the case for a ``classical'' state with macroscopic extension over
phase space. If one further rescales the dimensionless Wigner function to
$\tilde{W} \left( \Omega, {\boldsymbol{p}} \right) \equiv W \left(
\Omega, {\boldsymbol{m}} \right) / \left( 2 \hbar^3 \right)$ (where
$p_{\alpha}, p_{\beta}, p_{\gamma}$ replace $\hbar m_{\alpha}, 2 \hbar
m_{\beta} $ and $\hbar m_{\gamma}$), the quantum Liouville equation assumes
the form of the classical Liouville equation to leading order in $\hbar$,
\begin{eqnarray}
\fl
  \partial_t  \tilde{W} \left( \Omega, {\boldsymbol{p}} \right)  =&
   - \left[  \frac{1}{I_1 \sin^2 \beta} \left( p_\alpha - p_{\gamma}\cos \beta  
  \right) \frac{\partial}{\partial \alpha} + \frac{p_{\beta}}{I_1} 
  \frac{\partial}{\partial \beta} \right.
  \nonumber\\
  &+ \left\{ \frac{1}{I_1 \sin^2 \beta} \left(
   p_{\gamma} \cos^2 \beta - p_{\alpha}\cos \beta  \right) + \frac{p_{\gamma}
  }{I_3}  \right\}  \frac{\partial}{\partial \gamma}   
  \\
  &   \left. - \left\{ \frac{1}{I_1 \sin^3 \beta} \left( p_{\gamma} - p_{\alpha}\cos
  \beta  \right) \left( p_{\alpha} - p_{\gamma} \cos \beta  \right)
  \right\} \frac{\partial}{\partial p_{\beta}} \right] \tilde{W} \left(
  \Omega, {\boldsymbol{p}} \right) + \mathcal{O} \left( \hbar^2
  \right) . \nonumber
\end{eqnarray}
Hence, the classical dynamics is retained up to corrections of order
$\hbar^2$, like in the standard case of a point particle.

\begin{figure}[bt]
\begin{center}
  \resizebox{11cm}{!}{\includegraphics{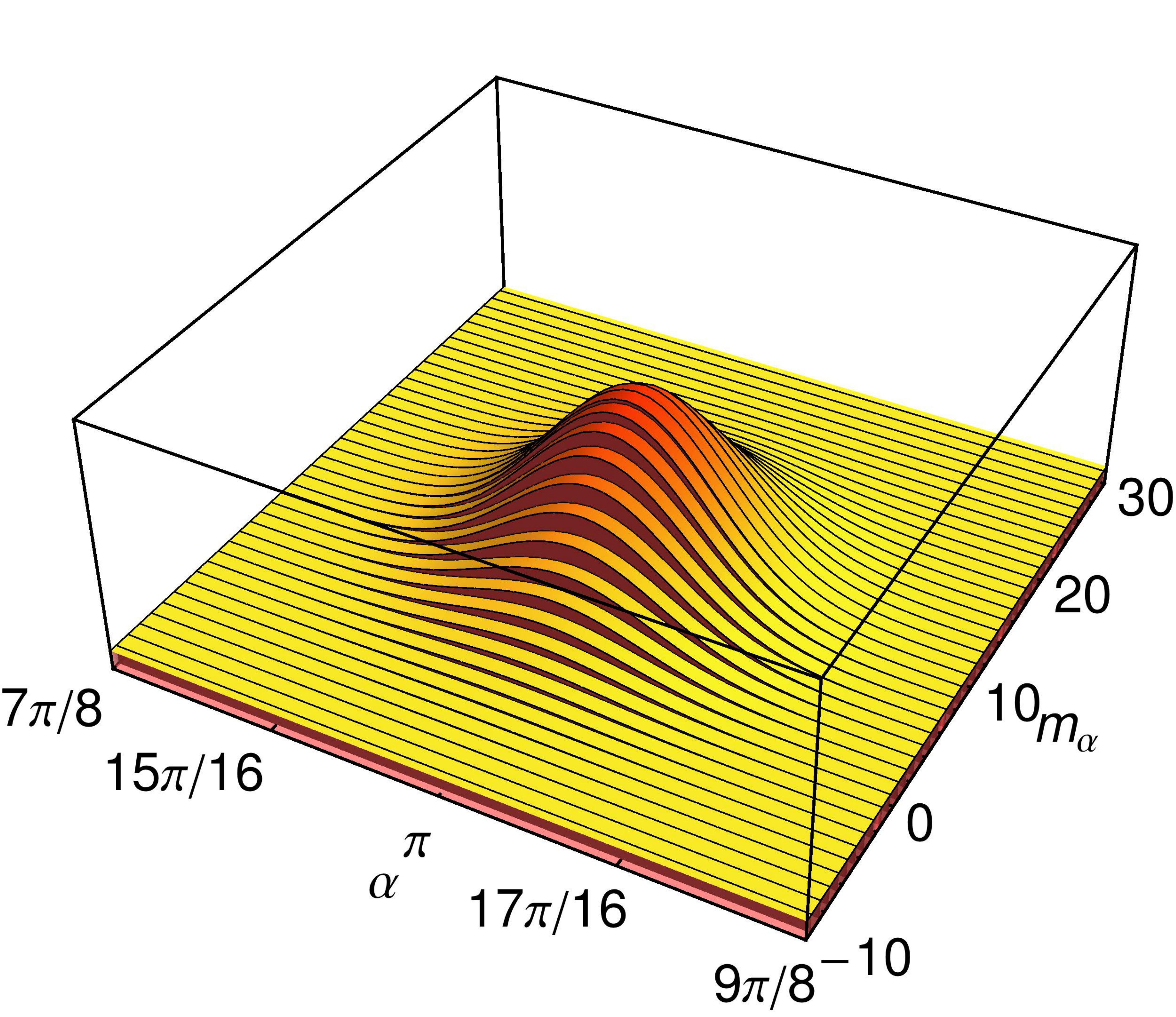}}
\end{center}
  \caption{Wigner function of a coherent state in $\alpha$ localized at
  $\left( \alpha, m_{\alpha} \right) = \left( \pi, 10 \right)$. The function is
  constant in the other angles and a Kronecker delta in the other angular
  momenta. \label{fig:coherent}}
\end{figure}

\section{Representative Wigner functions}
As a first application, we
consider the Wigner function of a coherent state in $\alpha$ {\cite{KK96}}, $|{\alpha, m_{\alpha}} \rangle_\alpha = \mathsf{D} \left( \alpha, m_{\alpha}
\right) | {0, 0} \rangle_\alpha  |m_{\beta} \rangle  |m_{\gamma}
\rangle$, where  
\begin{equation}
| {0, 0} \rangle_\alpha = \vartheta_3^{- 1 / 2} \left( 0,
{\rm e}^{ - 1 / \sigma^2 } \right) \sum_{m_{\alpha}} {\rm e}^{ - m_\alpha^2 /
\sigma^2 } |m_\alpha \rangle
\end{equation}
defines the coherent state at the phase space
origin (with $\vartheta_3$ a Jacobi theta function). Figure~\ref{fig:coherent}
shows the coherent state $| {\pi, 10} \rangle_\alpha$ with momentum spread $\sigma = 7$. As one expects, the Wigner function is well localized in
$\alpha$ and $m_{\alpha}$. If one decreases the angular width the discreteness
in $m_{\alpha}$ gets even less pronounced, illustrating that $m_{\alpha}$ can
be replaced by a continuous variable in the classical limit. We note that,
unlike the standard Wigner function for Glauber coherent states, the phase
space function of $| {\pi, 10} \rangle_\alpha$ takes on negative values in parts
of the phase space. However, since these are negligibly small compared to the
peak height in Fig.~\ref{fig:coherent} one can maintain that the coherent
states provide a classical correspondence. Coherent states involving more than
a single angle exhibit a similar behavior.

\begin{figure}[bpt]
  \begin{center}
\resizebox{10cm}{!}{\includegraphics{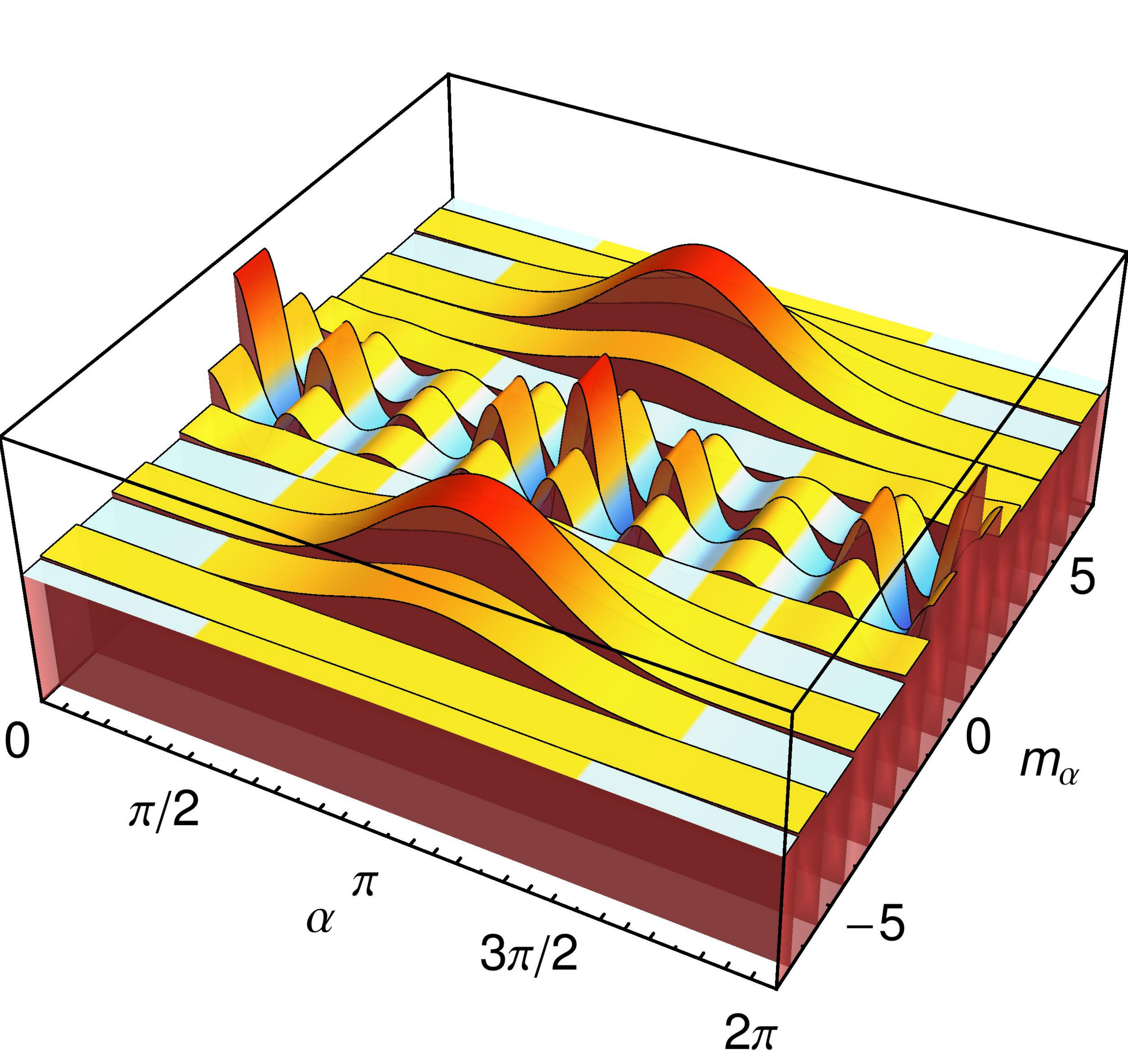}}
\end{center}
  \caption{Wigner function of a superposition of two $\alpha$-coherent states,
  localized at $\left( \alpha, m_{\alpha} \right) = \left( \pi, 4 \right)$ and
  $\left( \pi, - 4 \right)$. The number of fringes in the interference
  contribution is 8, as expected from $\Delta m_{\alpha} = 8$. The blue shaded
  colors indicate negative values ($\sigma = 1$).\label{fig:superpos}}
\end{figure}

If one superposes two coherent states separated in phase space, oscillatory
interference terms emerge in between the classical contributions, see
Fig.~\ref{fig:superpos}. We find that the number of interference fringes
scales with the phase space separation of the classical peaks. This is
precisely the behavior known from the standard Wigner function of superposed
Glauber coherent states.

\begin{figure}[bpt]
\begin{center}
  \resizebox{12cm}{!}{\includegraphics{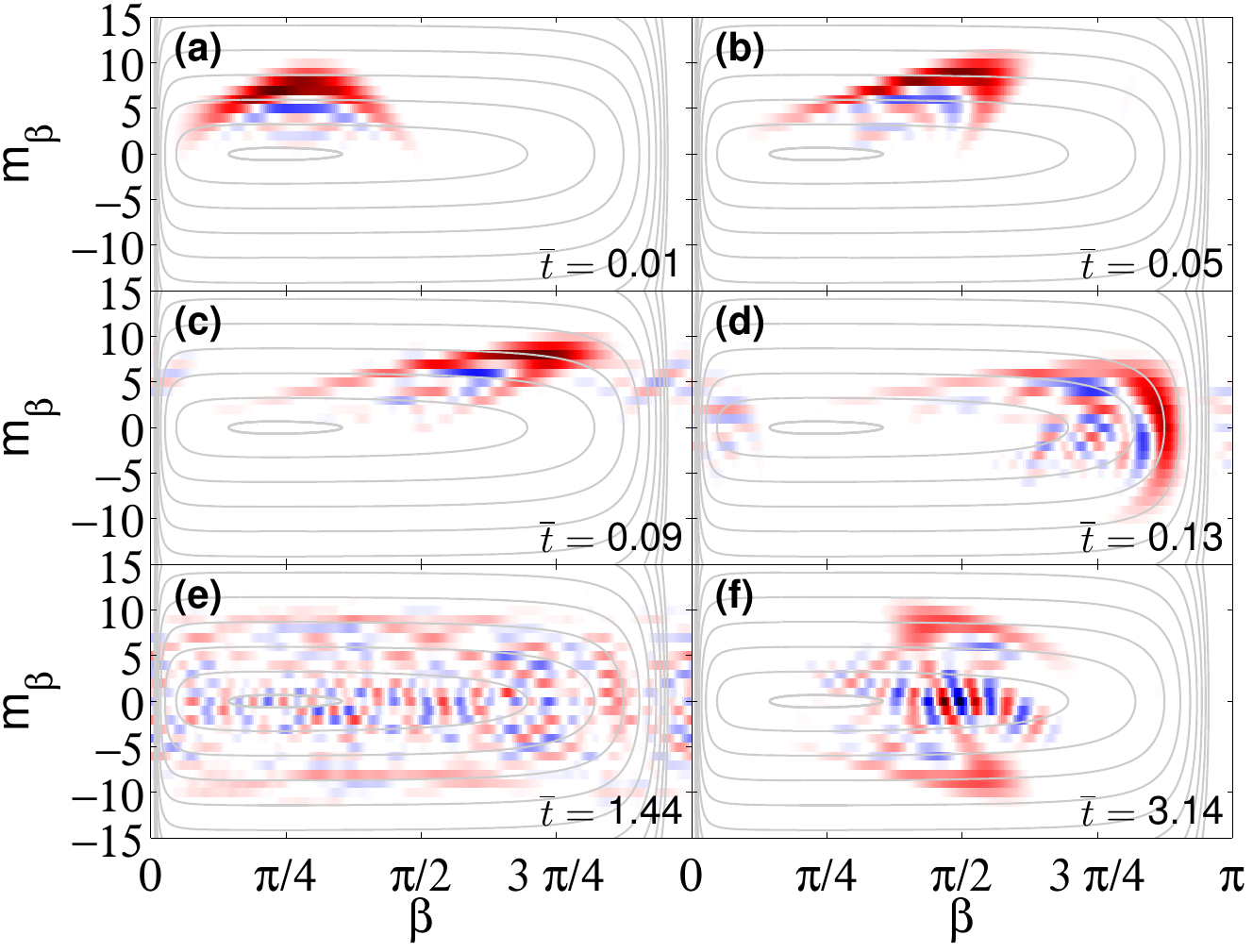}}
\end{center}
\caption{Phase space representation of molecular alignment dynamics. The red (blue) shades indicate positive (negative) values; gray lines represent classical phase space trajectories. Panel (a) depicts the state right after the kick, corresponding to maximal alignment. Panels (b), (c), and (d) show how the rotational wave packet evolves at the short times $\bar{t} = 0.05$, $0.09$ and $0.13$. After dispersion, at the later time $\bar{t} = 1.44$, it is completely delocalized and alignment is lost, see (e). A quantum interference effect related to the eigenvalue spacing of the symmetric top leads to an alignment revival at $\bar{t} = n \pi$, $n \in \mathbbm{N}$. Panel (f) provides the Wigner function at the first revival, while (a) is recovered for $\bar{t} = 2 \pi$. ($W \left( \Omega, {\boldsymbol{m}} \right)$ is constant in $\alpha, \gamma$ and a Kronecker delta in $m_{\alpha}, m_{\gamma}$.)
\label{fig:alignment}}
\end{figure}

\section{Molecular alignment dynamics}The
following application aims at demonstrating the viability of the presented
phase space approach also to describe dynamical situations. Specifically, we
study the nonadiabatic alignment of symmetric top molecules initiated by
picosecond laser pulses {\cite{TS05}}. In these experiments the expectation
value of $\cos^2 \hat{\beta}$ quantifies the alignment. The dynamics after the
initial laser kick is governed, in the rigid rotor limit, by the field-free
Hamiltonian of the symmetric top, $\hbar^2 \mathsf{H} = A \mathsf{J}^2 + \left(
C - A \right) \mathsf{J}_z^2$, where $A = \hbar^2 / 2 I_1$ and $C = \hbar^2 / 2
I_3$. Using the conventions of {\cite{TS05}}, we describe the initial kick by
a Gaussian pulse at $\bar{t} = 0$ with duration $10^{- 3}$ and strength
$\overline{\Omega_{}}_R = 10$ (time in units of $\hbar / I_1$).

Figure~\ref{fig:alignment} shows how the Wigner function of the molecular
orientation evolves in time, taking $|333 \rangle$ as initial state. The
half-moon shape in snapshot (a), shortly after the kick, is due to the
$\beta$-dependence of the interaction Hamiltonian
$\overline{\mathsf{H}}_{{\mathrm{int}}} = \bar{\Omega}_R 
\bar{\alpha}^{ZX} \cos^2 \hat{\beta}^{}$. The molecular alignment achieved can
be assessed from the form of the quasi-probability distribution. One also
notices an interference structure between the two arms of the half-moon,
indicating the coherence in the state. The following three snapshots (b), (c),
and (d) show the Wigner function at fractions of the period associated with
the classical trajectory centered on the half-moon. One observes that the
distribution function by and large follows the classical trajectories (gray
lines) at these short times, though it starts to disperse. In snapshot (e), at
a much later time, the distribution is spread over the accessible phase space
and the alignment is lost. Remarkably, at $\bar{t} = \pi$ (and in multiples of
thereof) the alignment revives, marked by a small dispersion in $\beta$, see
Fig.~\ref{fig:alignment}(f). One observes from the Wigner function that this phenomenon
can be associated with an angular momentum-superposition of two localized wave
packets.
A movie of the Wigner function dynamics is provided as supplementary material (available from http://stacks.iop.org/NJP/15/063004/mmedia).

\section{Conclusions}

In summary, a quantum phase space representation has been introduced for the rotation dynamics of rigid bodies. Based on the natural canonical variables, i.e. the Euler angles and the associated angular momentum projections, it inherits all the relevant properties of the standard Wigner function. This includes its interpretation as a quasi-probability distribution with correct marginals, its agreement with the Weyl correspondence rule and with the semiclassical limit. 
It provides an appropriate and intuitive framework for discussing quantum effects in the rotational motion, and we expect it to find applications in molecular physics \cite{stapelfeldt2003}, in quantum state reconstruction \cite{schmied2011}, and beyond.

This work was supported by the DFG (HO 2318/4-1).

\vspace*{\baselineskip}
%\bibliographystyle{iopart-num} 
%\bibliography{WignerBib}
\providecommand{\newblock}{}

\end{document}